\documentstyle[12pt]{article}
\newtheorem{lemmas}{Lemma}
\begin{document}
\parindent 8pt
\begin{titlepage}
\rightline{IASSNS-HEP-93/18}
\begin{center}
{\Large \bf  Exotic Black Holes?}
\end{center}\par
\medskip
\begin{center}
{\bf Carl H. Brans}
\end{center}\par
\medskip
\begin{center}

{Institute for Advanced Study\\ Princeton, NJ 08540\\ and\\
Physics Department\\ Loyola University\\ New Orleans, LA 70118  \\
e-mail:brans@loynovm.bitnet}\par\bigskip
\today
\end{center}
\begin{abstract}

Exotic smooth manifolds,  ${\bf R^2\times_\Theta S^2}$, are
constructed and  discussed as possible space-time models
 supporting the  usual Kruskal presentation of the vacuum Schwarzschild metric
locally, but {\em not  globally}.   While having the same topology as the
standard Kruskal model, none of these manifolds is diffeomorphic to standard
Kruskal, although under certain conditions some global smooth
Lorentz-signature metric can be continued from   the local Kruskal  form.
Consequently, it can be conjectured that such manifolds represent an infinity
of physically inequivalent (non-diffeomorphic) space-time models for
black holes.  However, at present nothing definitive can be said about the
continued satisfaction of the Einstein equations.
This problem is also discussed in the original Schwarzschild
 $(t,r)$ coordinates for
which the exotic region is contained in a world tube along the
time-axis, so that the manifold is spatially, but not temporally,
asymptotically standard.  In this form, it is tempting to speculate
that the confined exotic region might serve as a source for some
exterior solution.
Certain aspects of the Cauchy problem are also discussed in terms of ${\bf
R^4_\Theta}$\
models which are ``half-standard'', say for all $t<0,$ but for which
$t$ cannot be globally smooth.

\end{abstract}
\end{titlepage}

The process of extending investigations of space-time structures
from local to global  has clearly been of great value to theoretical
physics.  Standard physical theories use a smooth manifold model for
space-time, $M$, and define physical fields as cross sections of
bundles over this manifold.  As a smooth manifold $M$\ is
locally Euclidean, ${\bf R^4}$,\ in its topological and differentiable
properties, so expressions of theories as restrictions on cross
sections by the imposition of differential equations can always be
expressed locally as if space-time were standard ${\bf R^4}$.  However, from
the fairly early days of general relativity, it has become clear that the
global properties of $M$\ could have significant impact on physical
implications of a theory.  This progression has continued to the
present, from wormholes to topological defects.  Until fairly
recently it seemed that the only way to impose these non-trivial and
physically interesting global properties on $M$\ is through the use of
non-trivial topology.  Tacitly then, physics has been assuming that
different space-time manifold models which are {\em homeomorphic} are
necessarily also {\em diffeomorphic}.  Since diffeomorphisms are
commonly accepted as the mathematical representation of generalized
coordinate transformations, the fundamental principle of general
relativity then implies that such models (i.e. diffeomorphic ones) are
physically equivalent.  However, physics must now face the fact that
topologically identical manifolds need not be physically equivalent,
especially in the physically distinguished case of  dimension four.\par
In this paper,  the end-sum techniques of Gompf\cite{gem} are used to
construct and
 discuss manifolds, ${\bf R^2\times_\Theta S^2}$, which have the same topology,
${\bf
R^2}\times {\bf S^2}$, as the Kruskal space-time model, but which cannot be
diffeomorphic to it.  Thus,\par
{\parindent=0pt\bigskip {\bf Theorem:\ }{\em On some smooth manifolds which are
topologically
${\bf R^2\times S^2}$, the standard Kruskal metric
cannot be smoothly continued over the full range,
$u^2-v^2<1.$\bigskip\par}}\par
In order to provide the background for this result, let us begin
with a brief review of the relevant mathematical facts.  The
apparently innocuous question of whether or not  the set of differentiable
structures (modulo diffeomorphisms)
on ${\bf R^n}$ is trivial has long been of mathematical interest.  As
of about ten years ago, this question had been settled in the expected
affirmative for all $n\ne 4$, and probably most people expected the
exceptional case $n=4$ to ultimately resolve to the same conclusion.
After all, there is certainly no interesting topology in ${\bf R^4}$\ to
provide a basis for any other expectation. It was thus of considerable
interest when the existence of counter-examples began to appear around
1982, \cite{f},\cite{d},\cite{g0}.  Our paper
\cite{br} provides a brief survey of this problem and some conjectures
on the possible physical implications of these results.  In this paper,
certain questions raised in \cite{br} are at least partially answered.\par

Since the existence of non-trivial differentiable structures on
topologically trivial spaces is so strikingly counter-intuitive, it
is important to clarify several issues relating to
differential topology.  Specifically we must distinguish
the case of merely {\em
different} differentiable structures  from
{\em non-diffeomorphic} ones. The former are physically
indistinguishable, but the latter are definitely {\em not} physically
equivalent as
space-time models.  Consider a simple example,
$M\equiv{\bf R^1}$, which is to be regarded as a {\em
point set} with elements, $p$, which happen to be the real numbers.
 $M$\ is turned into a {\em topological manifold} by imposing the
standard real-number topology.  Next, for physics,
$M$\ must be supplied with a
{\em differentiable structure}.  This requires the definition of
a coordinate patch structure.  In this case, the standard structure
${\cal D}$,
can be defined over the entire manifold by global coordinates,
$x(p)=p$. That is, the coordinates are simply the numerical values
defining the point.\footnote{Technically ${\cal D}$\ is the maximal atlas of
all coordinate patch structures smoothly consistent with this one, but
this issue is not significant in this paper}.   $M$\ can be
endowed with many other ${\cal D}$'s.
For example, let $\bar{x}(p)=p^3.$  Clearly
this structure, ${\bar{\cal D}}$, is not consistent with the first one since
the map
$x\rightarrow \bar{x}=x^3$ is not smooth at the origin.  However,
the homeomorphism, $p\rightarrow p^{1/3}$ is actually a {\em
diffeomorphism}, $x\rightarrow \bar{x}=x$ when expressed in the local
smooth coordinates.  So, ${\cal D}$\ and ${\bar{\cal D}}$, while {\em
different} are
actually {\em diffeomorphic}. Since diffeomorphisms are regarded as generalized
coordinate transformations, this means that there is no new physics
available in ${\bar{\cal D}}$\  as compared to ${\cal D}$.  In fact, it is easy
to show
that ${\cal D}$\ on ${\bf R^1}$ is unique up to diffeomorphism, so any global
vs local
differences of physical significance can
only be obtained by topological alterations to
$M$, for example replacing ${\bf R^1}$ by ${\bf S^1}$. \par
However, precisely the opposite is the case for the surprising exotic
structures.  A smooth manifold  homeomorphic to ${\bf R^4}$\ but not
diffeomorphic to it is called ``exotic'' (or ``fake'') and denoted
here by ${\bf R^4_\Theta}$.\footnote{In general, the subscript $\Theta$ will
indicate a
non-standard object or process.  So $M\times_\Theta N$ means a smooth manifold
which is the topological, but not smooth,
cartesian product of the two manifolds. }\ Such a
manifold
 consists of a set of points which
can be globally topologically identified with the ordered set of four
numbers, say $(t,x,y,z)$.  While these may be smooth
coordinates locally over some neighborhood, they cannot be globally continued
as smooth functions.  Furthermore, in no diffeomorphic image of this
${\bf R^4_\Theta}$\ can the global topological coordinates be extended as
smooth
beyond some compact set.  This defining characteristic, which occurs
 only for dimension
four,  is in striking
contrast to the case of ${\bf R^1}$ discussed above.  There the
difference between the $p$ and the $p^3$ coordinates could be smoothed
away by a diffeomorphism.\par
Also, note that certain ${\bf R^4_\Theta}$\ have the property
that they
contain compact sets which cannot themselves be contained in the interior of
{\em any} smooth ${\bf S^3}$.  Thus, for some $R_0$, the {\em topological}
three-sphere,  $t^2+x^2+y^2+z^2=R^2$, cannot be {\em smooth} if $R>R_0$.
This is illustrated in Figure 1. Notice that in Figures 1
through 5 one space dimension has been suppressed, so each point is
actually a z-axis, while in Figure 6 two dimensions are suppressed and
each point is an ${\bf S^2}$.\par
  As interesting as these ${\bf R^4_\Theta}$\ are in their own right, a
technique
developed by Gompf\cite{g1} allows the construction of a large
 topological variety of exotic four-manifolds, some of which would
appear to have considerable potential for physics.
 Gompf's ``end-sum'' process  provides a
straightforward technique for constructing an exotic version, $M$, of any
non-compact four-manifold whose standard version, $M_0$, can be
smoothly embedded
in standard ${\bf R^4}$.  Recall that we want to construct $M$\ which is
homeomorphic
to $M_0$, but not diffeomorphic to it.  First construct a tubular
neighborhood, $T_0$, of a half ray in $M_0$. $T_0$ is thus standard ${\bf R^4}=
[0,\infty)\times {\bf R^3}$.  Now consider a diffeomorphism, $\phi_0$ of $T_0$
onto
$N_0=[0,1/2)\times{\bf R^3}$ which is the identity on the ${\bf R^3}$ fibers.
Do
the same thing for some exotic ${\bf R^4_\Theta}$\ with the important proviso
that it
{\em cannot} be smoothly embedded in standard ${\bf R^4}$.  Such manifolds are
known
in infinite abundance \cite{gem}.  Then construct a similar tubular
neighborhood for this ${\bf R^4_\Theta}$, $T_1$, with diffeomorphism, $\phi_1$,
taking
it onto $N_1=[1,1/2)\times{\bf R^3}$.  The desired exotic $M$\ is then
obtained by forming the identification manifold structure
\begin{equation}
M=M_0\cup_{\phi_0}([0,1]\times{\bf R^3})\cup_{\phi_1}\bf
R^4_\Theta\label{es1}\end{equation}
The techniques of forming tubular manifolds and defining
identification manifolds can be found in standard differential
topology texts, such as \cite{bj} or \cite{H}.  \par
Informally, what is being
done is that the tubular neighborhoods are being smoothly glued across
their ``ends'', each ${\bf R^3}$.  The proof that the resulting $M$\ is
indeed  exotic is then easy:  $M$\ contains ${\bf R^4_\Theta}$\ as a smooth
sub-manifold.
If $M$\ were diffeomorphic to $M_0$ then $M$,  and thus ${\bf R^4_\Theta}$,
could be
smoothly embedded in standard ${\bf R^4}$, contradicting the assumption on
${\bf R^4_\Theta}$.  Finally, it is clear that the constructed $M$\ is indeed
homeomorphic to the original $M_0$ since all that has been done
topologically is the extension of $T_0$. See figure 2 for a
visualization of this process when $M_0$ is ${\bf R^4}$.  Smoothly
``stuffing'' the upper ${\bf R^4_\Theta}$\ into the tube results in another
visualization of the new manifold as shown in figure 3.  A natural
doubling of this process leads to figure 4.  Finally, smoothly
spreading out the exotic tube in figure 3 leads to figure 5 \par
This is clearly a powerful technique for generating exotic manifolds.
Using it we are able to generate an infinity of non-diffeomorphic
manifolds, ${\bf R^2\times_\Theta S^2}$, each having the topology of the
Kruskal presentation
of the Schwarzschild metric.  Using the standard Kruskal notation
$\{(u,v,\omega); u^2-v^2<1, \omega\in S^2\}$ constitute global
{\em topological} coordinates, but {\em $(u,v)$ cannot be continued
as smooth functions over the entire range: $u^2-v^2<1$.}  However, by
techniques discussed in \cite{br}, these coordinates can be smooth over
some closed submanifold, say $A$, as illustrated in Figure 6.
Over $A$ then we can solve the vacuum
Einstein equations as usual to get the Kruskal form,
\begin{equation}
ds^2={32M^3e^{-r/2M}\over r}(-du^2+dv^2)+r^2d\Omega^2,\label{k1}\end{equation}
where $d\Omega^2$ is the standard spherical metric and
\begin{equation}
({r\over 2M}-1)(e^{r/2M})=v^2-u^2.\label{k2}\end{equation}
This metric form is thus valid over $A$, but cannot be extended beyond
it, not for any reasons associated with the development of
singularities in the coordinate expression of the metric, or for any
topological reasons, but simply because {\em the coordinates,
$(u,v,\omega)$, cannot be continued smoothly beyond some proper
subset, $A$,
of the full manifold},
thus establishing the theorem stated earlier.\par
However, given any Lorentzian metric on a closed
submanifold, $A$, some smooth continuation of the metric to all of $M$
can be guaranteed to exist under certain conditions.
  For example, we have
\begin{lemmas}
 If $M$ is any smooth connected 4-manifold and $A$ is a closed
submanifold for which
$H^4(M,A; {\bf Z})=0,$ then any smooth Lorentz signature metric defined over
$A$
can be smoothly continued to all of $M.$
\end{lemmas}

Proof: This is basically a question of the continuation of cross
sections on fiber bundles.  Standard obstruction theory is usually
done in the continuous category, but it has a natural extension to the
smooth class, as described on page 25 in \cite{steen}.
First, we note that
any Lorentz metric is decomposable into a Riemannian one, $g$,
plus a non-zero vector field,$v$.  The continuation of $g$ follows from
the fact that the fiber, $Y_S$, of non-degenerate symmetric four by four
matrices is $q$-connected for all $q$.  From the theorem stated on page
149 of \cite{steen} this means that $g$ can be continued from $A$ to all of $M$
without any topological restrictions. On the other hand, the fiber of
non-zero vector fields is the three-sphere which is $q$-connected for
all $q<3,$ but certainly not 3-connected ($\pi_3(S^3)={\bf Z}$).  From the
theorem and corollary on page 178 of \cite{steen}, any obstruction to a
continuation of $v$ from $A$ to all of $M$
is an element of $H^4(M,A;{\bf Z})$.  Thus, the vanishing of
this group is a sufficient condition for the continuation of $v$,
establishing the Lemma.
\par
In the applications in this paper, $M$ is non-compact, so
$H^4(M;{\bf Z})=0.$ Using the exact cohomology sequence generated by
the
inclusion $A\rightarrow M,$
\begin{equation}
  \cdots\rightarrow H^3(M;{\bf Z})
\rightarrow H^3(A;{\bf Z})\rightarrow H^4(M,A;{\bf Z})
\rightarrow H^4(M;{\bf Z})\rightarrow\cdots\label{exs1}\end{equation}
we see that one way to guarantee the condition of the Lemma is to
have $H^3(A;{\bf Z})=0.$ Another would be to establish that
the map, $H^3(M;{\bf Z})\rightarrow H^3(A;{\bf Z})$ is an epimorphism.
For  example, if $A$ is simply a closed miniature
version of ${\bf R^2\times S^2}$ itself, i.e.,
$A={\bf D^2\times S^2}$, then $H^3(A;{\bf Z})=0$ so
the continuation of a smooth Lorentzian metric is ensured.  Whatever
this metric is, it cannot be the Kruskal one, since otherwise the
manifold would be diffeomorphic to standard ${\bf R^2\times S^2}$.
An interesting variation of the situation described in
Figure 6 occurs when $A$  intersects the horizon. Thus it contains a trapped
surface, so a singularity will inevitably develop from well-known
theorems.  However, if $A$ does not contain a trapped surface
what will happen is
not known.
\par
What is missing from this result, of course, is that the continued metric
satisfy the vacuum Einstein equations and that it be complete in the
Lorentzian sense.  Of course, any smooth Lorentzian metric satisfies
the Einstein equation for some stress-energy tensor, but this tensor
must be shown to be physically acceptable.
 Unfortunately, these issues cannot be resolved
without more explicit information on the global exotic structure than
is presently available.  \par

Another way to study this metric is in terms of
 the original Schwarzschild
$(r,t)$ coordinates, as seen in figure 4.    For this model the coordinates
$(t,r,\omega)$ are smooth for all of the closed sub-manifold $A$ defined by
 $r\ge R_0>2M$ but cannot be continued
as smooth over the entire $M$ or over any diffeomorphic (physically
equivalent) copy.  In this case $A$ is topologically
${\bf R^1\times[}R_0{\bf ,\infty)\times S^2}$, so again $H^3(A;{\bf Z})=0$
and the conditions of lemma 1 are met.  Hence there is some
smooth continuation of any exterior Lorentzian metric in $A$, in particular,
 the Schwarzschild metric, over the full ${\bf R^4_\Theta}$.  Whatever this
metric
is, it cannot be Schwarzschild since the manifolds are not diffeomorphic.
  An interesting feature of this model is that the manifold is
``asymptotically'' standard in spite of the well known fact that
exotic manifolds are badly behaved ``at infinity''.  However, we note
that this model is asymptotically standard only as
$r\rightarrow\infty,$ but certainly not as $t\rightarrow\infty.$ \par
These models, especially as visualized in figures 3 and 4 are clearly
highly suggestive for investigation of alternative continuation of
exterior solutions into the tube near $r=0.$  We
often discover an {\em exterior}, vacuum solution, and look to continue
it back to some source.  This is a standard problem.  In the stationary
case, we typically have a local, exterior solution to an elliptic
problem, and try to continue it into origin but find we can't as a
vacuum solution unless we have a topology change (e.g., a wormhole),
or unless we add a matter source, changing the  equation.
Now, looking at figures 3 and 4, we are led to consider a third
alternative.\par
Of course, the discussion of stationary solutions involves the idea of
time foliations, which cannot exist globally for these exotic
manifolds, at least not into standard factors.
In fact,
\begin{lemmas}
${\bf R^4_\Theta}$\ cannot be written as a smooth product,
${\bf R^1\times_{smooth} R^3}$. Similarly
 ${\bf R^2\times_\Theta S^2}$\ cannot be written as ${\bf
R^1\times_{smooth}(
R^1\times S^2)}$.
\end{lemmas}
Clearly, if either factor decomposition were smooth, the original
manifold would be standard, since the factors are necessarily
standard from known lower dimensional results, establishing
the lemma.  I am indebted to  Robert Gompf and Duane Randall
 for pointing out to me that because of still open questions it is not
now possible to establish the
 more general result for which the second factor is simply some
smooth three manifold without restriction. Also, note that the question
of factor
decomposition of ${\bf R^4}$ into Whitehead spaces was considered by
McMillan\cite{mc}.\par
Of course, the lack of
a global time foliation of these manifolds means that such models are
 inconsistent with canonical approach to gravity,
quantum theory, etc.  However, it is worth noting that all
experiments yield only local data, so we have no {\em a priori} basis
for excluding such manifolds.\par
These discussions lead naturally to a consideration of what can be
said about Cauchy problems.
Consider then the manifold in figure 5.   The global $(t,x,y,z)$
coordinates are smooth for all $t<0$ but not globally. Now consider,
the Cauchy problem $R_{\alpha\beta}=0$, with flat initial
data on $t=-1$. This is guaranteed to have the complete flat metric
as solution in the standard, ${\bf R^4}$\ case.   However, the similar problem
{\em cannot} have a complete flat solution for ${\bf R^4_\Theta}$\ since then
the
exponential geodesic map would be a diffeomorphism of ${\bf R^4_\Theta}$\ onto
its
tangent space, which is standard ${\bf R^4}$.
This is discussed in \cite{br}.
What must go wrong in the exotic case, of course, is
that $t=-1$ is no longer a Cauchy surface.  However, Lemma 1 can again
be applied here to guarantee the continuation of {\em some} Lorentzian
metric over the full manifold since here $A=(-\infty,-1]\times{\bf R^3}$
so clearly $H^3(A;{\bf Z})=0$.  \par
Finally,  consider the cosmological
model, ${\bf R^1\times_\Theta S^3}$ discussed in \cite{br}.  In this case,
assume
a standard cosmological metric for some time, so here
$A=(-\infty,1]\times{\bf S^3}$.  Clearly,  $H^3(A;{\bf Z})$ does
not vanish in this case, but it can be shown that the inclusion
induced map
$H^3(M;{\bf Z})\rightarrow H^3(A;{\bf Z})$ is onto, so the conditions
of Lemma 1 are met. Thus  some smooth Lorentzian continuation will indeed
exist, leading to some exotic cosmology on ${\bf R^1\times S^3}$.
  \par
I am very grateful to Duane Randall and Robert Gompf for their
invaluable assistance in this work.
\par

\end{document}